\documentclass[12pt,twoside]{article}
\usepackage{fleqn,espcrc1}

\usepackage{graphicx}
\usepackage[figuresright]{rotating}

\newcommand{\be}{\begin{equation}}
\newcommand{\ee}{\end{equation}}  
\newcommand{\bea}{\begin{equation}} %eqnarray}}
\newcommand{\eea}{\end{equation}} %eqnarray}}

\newcommand{\AmS}{{\protect\the\textfont2
  A\kern-.1667em\lower.5ex\hbox{M}\kern-.125emS}}

\title{NT@UW-01-17  \\Light Front Nuclear Theory}

\author{Gerald A. Miller \address{Department of Physics, University of
  Washington,\\  
        Seattle, Washington 98195-1560 USA}
             \thanks{This work is partially supported by the USDOE. }} 
       
\begin{document}

% typeset front matter
\maketitle

\begin{abstract}
High energy scattering experiments involving nuclei are typically
analyzed in terms
of light front variables. The desire to provide realistic, relativistic
wave functions expressed in terms of these variables led me to try to
use  light front dynamics to compute nuclear wave functions.
Here calculations of infinite nuclear matter in the
mean field approximation and also in a light front version of Bruckner theory
which includes NN correlations are reviewed.
Applications of these
wave functions to nuclear deep inelastic scattering and Drell-Yan processes
are discussed. We find that relativistic mean field theory produces no
EMC binding effect.
\end{abstract}

\section{INTRODUCTION}

The goal of our approach\cite{Miller:1997xh,Miller:1997cr,Miller:2000kv} 
is to test the assumption that the nucleus is made of nucleons and mesons.
Thus we attempt to compute the full Fock state wave function of such a system
using light front dynamics.

We shall follow the following outline, beginning with trying to answer the
questions ``What is light front nuclear theory?'' and more importantly,
``Why do it''? I will keep the discussion of the formalism to a minimum,
as the details have been published. 
Then we proceed to examples and results. First,  the properties of infinite
nuclear matter, obtained using the mean field approximation 
will be discussed.
The application of this theory to finite nuclei has been
made\cite{Blunden:1999hy},\cite{Blunden:1999gq},
 but will not be discussed here.
Then nucleon-nucleon correlations are included by
implementing a light front version of Bruckner theory.

A general result of these considerations is that vector and scalar mesons
are prominent components of
nuclear wave functions. This is important in trying to explain the
HERMES effect\cite{Ackerstaff:2000ac,Miller:2000ta}.
We also find that pionic effects are small, but testable. 
This talk is based on work with  collaborators P. Blunden, M.Burkardt,
J. Cooke, R. Machleidt, J. Smith and B. Tiburzi.
                                
\section{ WHAT IS LIGHT  FRONT  DYNAMICS?}

This is a relativistic many-body dynamics in which fields are quantized at a
``time''=$t+z=x^0+x^3\equiv x^+$. The canonical energy is then given by
$P^0-P^3\equiv P^-$. These equations show the  notation that    
a four-vector $A^\mu$ is expressed in terms of its $\pm$ components
\be A^\pm\equiv A^0\pm A^3.\ee
The operator $P^-$ acts as an $x^+$ evolution operator. 
One quantizes at $x^+=0$ which is a light-front, hence the name ``light front
dynamics''.

 The canonical spatial variable must be orthogonal to the time variable,
 and this is given by 
$x^-=x^0-x^3$. The canonical momentum is then $P^+=P^0+P^3$. The other
coordinates are as usual ${\bf x}_\perp$ and  ${\bf P}_\perp$.

A consequence of our notation is that  dot products are written as
\be A\cdot b=A^\mu B_\mu={1\over2}(A^+B^-+A^-B^+)-{\bf A}_\perp \cdot
{\bf B}_\perp.\ee The most important result is the relation between energy and
momentum:
\be p_\mu p^\mu=m^2=p^+p^--p_\perp^2,\ee
which means \be p^-={p_\perp^2+m^2\over p^+}.\ee
This is a relativistic formula for the kinetic energy which does not contain
a square root operator. This is very useful in separating the center of mass
and relative coordinates, so that the computed wave functions are frame
independent. To a first approximation one may say that doing light front
dynamics is doing ordinary quantum mechanics, but with energy denominators
obtained using $p^-$. 
\section          {MOTIVATION FOR  LIGHT FRONT NUCLEAR PHYSICS}
Much of this
work is
motivated by the desire to understand nuclear deep inelastic scattering
and related experiments, so
it is worthwhile to review some of the features of the EMC
effect \cite{emc,emcrevs,fs2}. One central  experimental result  is the
suppression of the structure function for $x\sim 0.5$. The immediate
parton model interpretation is    that
the valence quarks of bound nucleons carry less plus-momentum than
those of free nucleons. Some other degrees of freedom are therefore needed
to maintain to total  plus-momentum, and some authors therefore % This may be understood by
postulate that mesons carry a larger fraction of the plus-momentum in
the nucleus than in free space\cite{chls,et}.
While such a model explains the shift
in the valence distribution, any meson contains valence 
anti-quarks and that distribution is  enhanced
compared to free nucleons. This  should be observable in Drell-Yan
experiments \cite{dyth}, but  no such enhancement has been observed
experimentally \cite{dyexp}, and  this has been termed as a severe crisis
for nuclear theory in
Ref.~\cite{missing}.

The EMC effect is rather small, so that one may begin by regarding the
nucleus as being made of nucleons. In this case, we say that 
deep inelastic scattering proceeds when a virtual photon is absorbed by
a quark carrying plus-momentum $p^+$, which came from a nucleon carrying
a plus-momentum $k^+$. In the parton model, the kinematic variable
$x_{Bj}=Q^2/2M_N\nu $ is given by 
\be
x_{Bj}={p^+\over P^+}={p^+\over k^+}{k^+\over P^+},\ee
where $P^+$ is the momentum of the nucleus $P^+=M_A$ in the target rest frame.
Thus one needs to know
 the                 probability $f_N(k^+/P^+)$
that a nucleon has a momentum fraction $k^+/P^+$. One
also wants to know the related probability for a
meson, for example: $f_\pi(k^+/P^+)$.

The essential technical advantage of using light cone variables
is that the light cone energy $P^-$ of a given final state 
does not appear in the delta function which expresses the
conservation of energy and momentum. Thus one may use closure to
perform the sum over final states which appears in the calculation of 
an exclusive nuclear cross section. The result is that the
conventional lore is that
cross sections may be expressed in terms of the probabilities:
\be
\sigma\propto f_N({k^+\over P^+})\sim\int d^2k_\perp\cdots
\mid \Psi_{A}({k^+\over P^+},{\bf k}_\perp,\cdots)\mid^2,\ee where
$\Psi_A$ represents the ground state wave function.
Another less formal way of saying the same thing is that two events
at ($z_1,t_t)$ and ($z_2,t_2)$ which have a light-like separation occur at
different times, but at the same value of $x^+=z_1+t_1=z_2+t_2$. Thus one
avoids the need to use a time development operator which would lead to summing
over an infinite number of excited states. Such a light-like separation occurs
often when one considers high energy scattering processes.

For  these reasons we are concerned  with        calculating
the distribution function
$f_N(k^+/P^+)$. Since usual nuclear dynamics is done within the equal-time
 formulation, the $k^+$ variable is not readily available.  Thus we need 
  realistic calculations, with real dynamics and
symmetries. This brings me to the conclusion 
 that it is necessary                  to
 redo nuclear physics on the light front.
The main motivation here is to do a good calculation of conventional dynamics.
If the calculations are good enough and fail to reproduce data then one may be
able to reach an interesting conclusion that interesting non-standard dynamics
are involved. 
 
\section{LIGHT FRONT QUANTIZATION LITE}
You have to have a Lagrangian  ${\cal L}$   no matter how bad!
This is because, in contrast with approaches based on symmetries,
we try to obtain all of the necessary operators from a given 
Lagrangian ${\cal L}$.
We use
%\cite{Miller:1997xh,Miller:1998tp}
% put your own definitions here:
  ones in which the degrees of freedom are 
nucleons, vector and scalar mesons and pions. The  existence of ${\cal L}$ 
allows the  derivation of
the canonical, symmetric energy momentum tensor $T^{\mu\nu}$.
In light front dynamics the momentum is $P^+=P^0+P^3$ where $P^\mu$
is the total
momentum
operator: $P^\mu={1\over 2}\int d^2x_\perp dx^-\;T^{+\mu}$.
 One necessary detail is that
that $T^{+-}$ must be expressed in terms of independent degrees of freedom.
One uses the equations of motion to express the dependent degrees of freedom
in terms of the independent ones, and uses these constraint equations in the
expression for $T^{+-}$.  

We use two Lagrangians. The first is that of the  Walecka model\cite{bsjdw}:
  ${\cal L}(\phi,V^\mu,N)$ which contains the fields:
 nucleon $N$, neutral vector meson  $V^\mu$, neutral scalar meson $\phi$.
This is the simplest model which provides a reasonable caricature  of
the nucleus. The binding is caused by the attractive effects occurring at 
relatively long range 
when nucleons exchange scalar mesons. The nucleus  is prevented from collapsing
by the short distance repulsion arising from the exchange of vector mesons.

We also shall show results obtained using a more complicated chiral Lagrangian:
in which the fields are $  N,\pi,\sigma,\omega,\rho,\eta,\delta$. Our plan
is to  first use the 
     Walecka model in the       mean field approximation, and then 
to  employ a chiral Lagrangian and include nucleon-nucleon correlations
in a formalism which yields a non-zero pionic content.
\section { INFINITE NUCLEAR MATTER IN  MEAN FIELD APPROXIMATION-- WALECKA
MODEL }

In this approach one assumes that
the strong baryon sources  produce   sufficiently many mesons to justify a
classical    treatment.          In infinite nuclear matter, one works in
a limit in which   the 
nuclear volume  $\Omega$ is considered to be  infinite, so that 
 all positions, and directions  equivalent in the nuclear rest frame.
Then  the 
fields $\phi$ and $V^\pm$ are
constant, with    ${\bf V}_\perp=0$. These features
simplify the solution of the 
           field equations. One
easily obtains the operators $
T^\pm$, and the light front ``momentum'' and ``energy'' are given by 
\be {P^\pm\over\Omega}=\langle
T^\pm\rangle,\ee in which
 the expectation value is over  the nuclear ground state.

%\smallskip Self-consistent solution!

The  nuclear momentum content is the essential feature we
 wish to understand here. The results are that  \be
{P^-\over\Omega}=m_s^2\phi^2+{4\over
(2\pi)^3}\int_F d^2k_\perp dk^+\;{k_\perp^2+(M+g_s\phi)^2\over k^+},\ee
\be{P^+\over\Omega}=m_v^2(V^-)^2+{4\over
(2\pi)^3}\int_F d^2k_\perp dk^+\;k^+.\label{p+}\ee
The first term of $P^+$ is  the plus momentum carried by vector mesons,
and the second term is the plus momenta carried by the nucleons.
Here $g_s$ is the scalar-meson-nucleon 
coupling constant, and the vector meson-nucleon coupling constant 
$g_v$ enters in the expression for $V^-$.
The interpretation of these results is aided by a change of variables:
\be
k^+\equiv \sqrt{(M+g_s\phi)^2+\vec{k}^2} +k^3,\label{k3}\ee  which
defines defines the variable $ k^3$. Using this variable one can show that 
rotational invariance is respected and obtain a spherical Fermi surface. 
The Fermi sphere is defined via Eq.~(\ref{k3}) by the
relation
\be k_\perp^2+(k^+-E_F^*)^2\le k_F^2,\quad E_F^*\equiv
\sqrt{k_F^2+{M^*}^2}.\label{fermisphere} \ee
Furthermore, one may show that  
$E\equiv{1\over 2}\left(P^-+P^+\right)$ is the           same   as
the usual expression obtained in the  Walecka  model.

  For nuclear matter in its rest frame we need to obtain
 $P^+=P^-=M_A$. This is the light front expression  of the statement that
the pressure on the system must vanish\cite{Miller:1998tp,Miller:1999ap}. Indeed 
 the minimization  
\bea\left({\partial (E/A)\over\partial k_F}\right)_\Omega=0,\eea
determines the value of the Fermi momentum and is an expression
that gives %sets %\to\;k_F,\quad
$P^+=P^-=M_A$.  

We can quickly obtain the relevant numerical results using the 1974
parameters of 
 Chin \& Walecka. These are 
\bea g_v^2M_N^2/m_v^2=195.9\qquad g_s^2M_N^2/m_s^2=267.1.\eea
With these parameters 
$M_N+g_s\phi =0.56 M_N$ and $g_vV^-=270$ MeV. These are the  HUGE
scalar and vector potentials which are characteristic of the Walecka
 model.
The interesting variables are those associated with 
the  total nuclear plus momentum
 ${P^+\over \Omega}$. With  the above parameters, the 
vector meson contribution to this quantity: $%{P_V\over \Omega}=
m_v^2(V^-)^2$ is a monumental $0.35\;{P^+\over \Omega},$ while the nucleon
contribution  $
{4\over
(2\pi)^3}\int_F d^2k_\perp dk^+\;k^+$ is only 
$0.65{P^+\over\Omega}$. 
Only 65 \% of $P^+$ carried by nucleons,
but 90\% is needed to understand the EMC effect in infinite nuclear
matter\cite{Sick:1992pw}.

\section{  PLUS-MOMENTUM
 DISTRIBUTIONS}
The nucleonic contribution to the nuclear
plus momentum is given by  the relation (\ref{p+}),so %  component of the 
 the probability $f(k^+)$ that a nucleon carries a plus momentum 
$k^+$  may be  defined by: % given by the relation
%\begin{equation}
  $P^+_N/A\equiv\int
dk^+\;k^+ f(k^+),
$ %\end{equation}
with
\bea
f(k^+)={4\over\rho_B (2\pi)^3}\int_F d^2k_\perp=
{4\over\rho_B (2\pi)^3}\int_F d^2k_\perp dp^+\delta(k^+-p^+),\label{fdef}
\eea 
where the subscript $F$ denotes integration over the Fermi sea according to
Eq.~(\ref{fermisphere}). It is useful to obtain a dimensionless distribution
function $f(y)$  by 
replacing $k^+$ by the dimensionless variable $y$ using
$y\equiv {k^+\over \overline{M}},f(y)\equiv\overline{M}f(y\overline{M}),$
with $\overline{M}\equiv P^+_A/A=M_A/M$
Then using Eq.(\ref{fermisphere})
leads to the result
\begin{equation}
f(y)={3\over 4} {\overline{M}^3\over k_F^3}\theta(y^+-y)\theta(y-y^-)\left[
{k_F^2\over \overline{M}^2}-({E_F^*\over \overline{M}}-y)^2\right], \label{fy}
\end{equation}
where
$y^\pm\equiv {E_F^*\pm k_F\over \overline{M}}$ and
$E_F^*\equiv\sqrt{k_F^2+{M^*}^2}$. %+g_s\phi)^2}$. 
This function peaks narrowly at the value $y=0.65$, which causes a disaster
if the conventional lore is used to compute the nuclear structure function
$F_{2A}$:
\begin{equation}
{F_{2A}^{\rm lore}(x)\over A}=\int dy f(y) F_{2N}(x/y). \label{deep}
\end{equation}
There is far, far too large a depletion of $F_{2A}$ because here the nucleons
carry only 65\% of the plus-momentum.
One can't plot the results of the theory in
comparison with the experimental results using pages of ordinary size.

However, Eq.~(\ref{deep}) is obtained without having made
the connection between the nucleon momentum distribution
computed using light front dynamics and that used in computing the deep
inelastic structure function. Recently, the authors of
Ref.~\cite{HB} have claimed that quark distribution functions are not parton
probabilities. The key point is that
one needs to derive the connection between the constituent distribution
function
and the observed data. That work stimulated us to undertake an
investigation\cite{mreturn} in which this connection is derived.
That investigation uses a manifestly covariant formulation\cite{jm} and the
result is a formula
\begin{equation}
{F_{2A}^{\rm lore}(x)\over A}=\int dy f(y-{V^+\over\overline{M}})
F_{2N}(x/y), \label{deeper}
\end{equation}
so that now one using a nucleon distribution function which has  peak at unity.
This leads to no depletion. There is only an enhancement due to Fermi motion.
This result is obtained for any relativistic mean field theory.
A very similar result was obtained by Birse some time ago\cite{birse}.
A paper on this subject has recently been submitted for
publication\cite{mreturn}.

\section{  BEYOND MEAN FIELD THEORY} %beyond  Mean Field Theory}

The interactions between nucleons are strong, and the mean field approximation
is unlikely to provide a description of nuclear properties which involve
high momentum observables.  We developed
\cite{Miller:1997cr,Miller:1998tp,Miller:1999ap}
a version of light
front theory in which the correlations between two nucleons are included.
The theory was applied to infinite  nuclear matter.

The calculation required three principal steps. (1)
 Light front quantization of
chiral ${\cal L}$. (2)
 Derive Light Front version of the $NN$ one boson
exchange potential. This could be done exploiting the relationship
between the Weinberg equation and the Blankenbecler-Sugar equation.
Results for the phase shifts are shown  in Fig.~1 of
Ref.~\cite{Miller:1999ap}. Generally good agreement with the measured data
is obtained. The light front approach is no worse than any other. 

The third step is to develop the many body theory. This turns out to be
a long story\cite{Miller:1999ap}. %, but the net result is that the formalism  looks
%just like ordinary non-relativistic many body formalism.
The net result is that the light front theory looks like the usual relativistic
Brueckner theory 
theory except that the Blankenbecler-Sugar equation is used, and the effects
of retardation are kept. The resulting nuclear matter saturation curve is
shown  in Fig.~2 of Ref.~\cite{Miller:1999ap}.
 Standard good results for nuclear
saturation properties are obtained, with a possible improvement in the lowered
 value, 180 MeV, of the  computed nuclear compressibility. One may compare
 the present  calculations with those of
 earlier relativistic Bruckner theory calculations, e.g. that of
Ref.~\cite{Brockmann:1990cn}, by comparing  the light front equations with
those of the usual equal time formulation. The differences reduce to the use of
teh Blankenbecler-Sugar equation instead of the Thompson equation, and to the
inclusion of retardation effects
(necessary for Poincare invariance) in the light front approach.

 The results for deep inelastic scattering and the related Drell-Yan process
 seem very promising.
Our preliminary result\cite{we} is that
the nucleons carry about  93\% of the nuclear plus
 momentum, which includes the effects of 
 two-particle two-hole states. The
 calculation of lepton-nucleus deep inelastic scattering is not finished.
We are working on it\cite{we}

In these calculations the nucleus
 does have a pionic component, which arises as a result of going beyond the
 mean field approximation. 
 The number of excess pions per nucleon was computed to be
 about 5\%\cite{Miller:1999ap}, which accounts for about 
 2\% of the nuclear plus momentum\cite{we}, 
  small enough to avoid a  contradiction with the Drell-Yan
data\cite{jm,reveal}. This 5\% is smaller than the value of
$\sim$20\% found in Ref.~\cite{bf}, but is in accord with the more recent
calculation of Ref.~\cite{Akmal:1997ft}. The difference between the
calculations
is related to the inclusion of the $\Delta$ as an explicit degree of freedom.
This was done in Ref.~\cite{bf}, but not in our or the more recent Illinois
calculations.

It may be possible to observe  the pionic content by
measuring an enhancement of the cross sections $\sigma_L$
for longitudinally polarized
photons in electron nucleus scattering\cite{reveal}. There are 15-30\%
enhancement effects at values of $x\sim0.2-0.3$ (shown in Fig.~4 of
Ref.~\cite{reveal})
which present an excellent
opportunity to unravel along-standing mystery about the presence or absence of
pionic effects.

\section{SUMMARY} 

The light front approach leads to  boos-invariant wave functions of the form
$\Psi_A({k^+\over P^+}$. Furthermore, formalisms to compute these wave functions
do exist right now.
Mesons are significant components of nuclei, and can be used to explain the
HERMES effect. Relativistic mean field theory can not reproduce the EMC effect
of nuclear depletion of the structure function for $x\ge  0.4$. One may go
beyond
mean field theory and include the correlations. This leads to a small pionic
effect, which causes inconsistency with the nuclear Drell-Yan experiment, but
which is testable by measuring $\sigma_L$ in
electron nucleus scattering.

\end{document}